\pdfoutput=1
\documentclass[11pt, a4paper]{article}

\usepackage[british]{babel}
\usepackage{hyperref}
\usepackage{siunitx}
\usepackage{upgreek}
\usepackage{textgreek}
\usepackage{subfigure}
\usepackage{booktabs}
\usepackage{amsmath}
\usepackage{authblk}
\usepackage{url}
\usepackage{graphicx}
\usepackage[labelfont=bf, font=small]{caption}

\newcommand{\katrin}{\textsc{Katrin}}
\newcommand{\tritiummol}{T$_2$}

\hypersetup{
	colorlinks=true,	% if true -> remove boxes around links
	urlcolor=blue,
	linkcolor=black,
	citecolor=blue,
	pdfauthor={Florian Heizmann},
	pdfsubject={Not set},
	pdfkeywords={Not set}
}

\title{Modelling of gas dynamical properties of the {\normalfont\katrin} tritium source and implications for the neutrino mass measurement}
\author[a,*]{Laura~Kuckert}
\author[b,*]{Florian~Heizmann}
\author[b]{Guido~Drexlin}
\author[a]{Ferenc~Gl\"uck}
\author[b]{Markus~H\"otzel}
\author[b]{Marco~Kleesiek}
\author[c]{Felix~Sharipov}
\author[a]{Kathrin~Valerius}

\affil[a]{Institute for Nuclear Physics, Karlsruhe Institute of Technology (KIT), Hermann-von-Helmholtz-Platz~1, 76344~Eggenstein-Leopoldshafen, Germany}
\affil[b]{Institute of Experimental Particle Physics, Karlsruhe Institute of Technology (KIT),  Wolfgang-Gaede-Str.~1, 76131~Karlsruhe, Germany}
\affil[c]{Departamento de F\'{\i}sica, Universidade Federal do Paran\'{a}, Caixa~Postal~19044, 81531-980, Curitiba, Brazil}
\affil[*]{Corresponding authors: Laura~Kuckert, \href{mailto:laura.neumann@ewetel.net}{ {\normalfont laura.neumann@ewetel.net} } and Florian~Heizmann, \href{mailto:florian.heizmann@kit.edu}{ {\normalfont florian.heizmann@kit.edu} } }

\date{\vspace{-5ex}}

\begin{document}
\maketitle

\begin{abstract}
	The \katrin~experiment aims to measure the effective mass of the electron antineutrino from the analysis of electron spectra stemming from the \textbeta-decay of molecular tritium with a sensitivity of \SI{200}{\milli\electronvolt\per c\squared}.
	Therefore, a cumulative amount of about \SI{40}{g} of gaseous tritium is circulated daily in a windowless source section. 
	An accurate description of the gas flow through this section is of fundamental importance for the neutrino mass measurement as it significantly influences the generation and transport of \textbeta-decay electrons through the experimental setup. In this paper we present a comprehensive model consisting of calculations of rarefied gas flow through the different components of the source section ranging from viscous to free molecular flow. By connecting these simulations with a number of experimentally determined operational parameters the gas model can be refreshed regularly according to the measured operating conditions. In this work, measurement and modelling uncertainties are quantified with regard to their implications for the neutrino mass measurement. 
	We find that the magnitude of systematic uncertainties related to the source model is represented by $\left|\Delta m_\upnu^2\right|=\left(3.06\pm 0.24\right)\cdot\SI{e-3}{\electronvolt\squared}/\textrm{c}^4$, and that the gas model is ready to be used in the analysis of upcoming \katrin~data.
\end{abstract}


\tableofcontents

% include (with clearpage at begin/end) vs input (no clearpage)
\section{Introduction}\label{sec:introduction}
\begin{figure}[t]
	\includegraphics[width=1.0\textwidth]{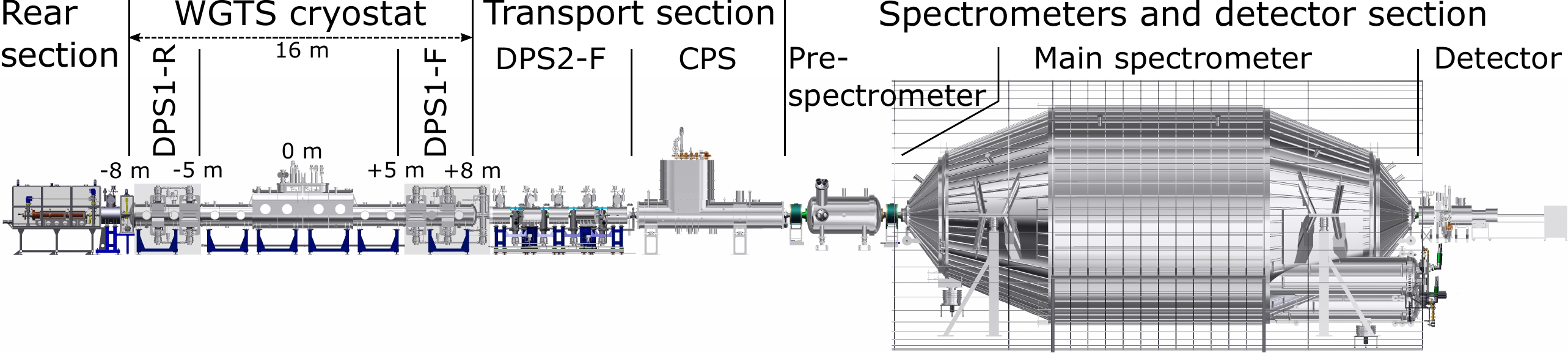}
	\caption[Overview of the \katrin~experiment.]{Overview of the \katrin~experiment. Tritium gas is injected in the source (WGTS) and pumped out in adjacent pumping sections (DPS1/2, CPS). Electrons from \textbeta-decay are magnetically guided to the energy analysing spectrometer section and are counted at the detector.}
	\label{fig:katrin_overview}
\end{figure}
The determination of the absolute mass scale of neutrinos is one of the most fundamental open challenges in particle physics. A model-independent determination in a laboratory experiment can only be provided by experiments using the kinematics of \textbeta-decay like the KArlsruhe TRItium Neutrino (\katrin) experiment which is currently in its commissioning phase.

\katrin~is designed to reach an unprecedented neutrino mass sensitivity of \SI{200}{\milli\electronvolt} (90\%\,C.L.) by high-precision tritium \textbeta-decay spectroscopy combined with an ultra-luminous gaseous tritium source~\cite{KATRIN2004}. A schematic overview of the \katrin~experiment is shown in fig.~\ref{fig:katrin_overview}. The  Windowless Gaseous Tritium Source (WGTS)~\cite{Grohmann2008,Babutzka2012,Priester2015,Heizmann2017} will provide a large \textbeta-decay rate of \SI{e11}{\per\second} by circulating a daily throughput of \SI{40}{g} of tritium, resulting in a column density in the beam tube of $\mathcal{N}=\SI{5e21}{\per\square\metre}$. 

To prevent tritium from migrating into the spectrometers, the gas flow needs to be reduced by 14 orders of magnitude in adjacent pumping sections  by kinetic (differential pumping section, DPS1/2) and cryogenic (cryogenic pumping section, CPS) pumping. The pre- and main spectrometer are of MAC-E filter type~\cite{Beamson1980,Kruit1983,Lobashev1985,Picard1992} and allow high-resolution energy analysis of the \textbeta-decay electrons by scanning the electrostatic spectrometer retarding potential. 

The neutrino mass will be extracted by comparison of the experimentally measured electron spectrum to a theoretically modelled equivalent~\cite{Kleesiek2014,Kleesiek:2018mel}. The modelling takes into account a variety of experimental effects, among which the electron-gas inelastic scattering processes inside the WGTS are of particular importance as they modify the electron energy. Understanding this effect requires precise knowledge of the column density $\mathcal{N}$, or the number density of gas molecules integrated along the beam tube axis, which is also an important input for plasma simulations.

In addition, the knowledge of the axial gas density distribution in the source section is necessary to correct for spatial inhomogeneities of parameters influencing the electron spectrum, such as magnetic field and temperature. The gas dynamics model used to determine this density distribution needs to cover a broad range of pressure regimes while providing a total uncertainty of \num{0.2}\% on the product of column density and scattering cross section, $\mathcal{N}\cdot\sigma$. At the same time, the gas dynamics model must be adjustable to account for varying operational parameters such as temperature and inlet pressure.

Malyshev et al.~\cite{Malyshev2009} described parts of such a highly accurate gas model focussing on the calculation of gas flow reduction factors. However, due to changes in the apparatus, these calculations need to be updated and refined. 
The goals of this work are to i) describe this refined gas model and ii) analyse its impact on the neutrino mass measurement considering experimentally determined parameters.
 
We start in section~\ref{sec:wgts} with a description of the source and how its \textbeta-decay electron spectrum can be modelled before introducing the gas dynamics calculations of the particular components in section~\ref{sec:gasdynamics-theory-and-calculations}. Moreover, section~\ref{sec:source-modeling} presents the determination of the column density by combining measurement and calculation. The corresponding uncertainties are analysed and their impact on the neutrino mass measurement is investigated. Finally, in section~\ref{sec:conclusion}, we  conclude this work with a summary of our findings.

\section{Electrons from the {\normalfont\katrin}~source section}\label{sec:wgts}
The Windowless Gaseous Tritium Source (WGTS) provides \textbeta-decay electrons via a continuous tritium throughput of \SI{1.8}{\milli\bar\litre\per\second}  (related to a temperature of \SI{273.15}{\kelvin}). 
\num{99}\% of the decays happen in the central beam tube with a length $L$ of \SI{10}{\metre} and a diameter \O~of \SI{90}{\milli\metre} to which differential pumping sections are attached at both the front and rear ends (DPS1-F and DPS1-R, see fig.~\ref{fig:katrin_overview}). Those turbomolecular pumps of type \textit{Leybold TURBOVAC MAG W 2800} have a pumping speed of about \SI{2000}{\litre\per\second} each.

The beam tube is surrounded by superconducting magnets that produce a homogeneous and stable magnetic field of \SI{3.6}{\tesla}, all housed within a complex large-scale cryostat infrastructure.
The beam tube wall temperature is stabilised at \SI{30}{\kelvin} to better than \num{0.1}\% using a two-phase neon cooling system \cite{Grohmann2009} based on two coolant pipes lining the beam tube. A proof of concept of the high stability cooling system was performed with a \textit{Demonstrator} set-up \cite{Grohmann2013} and has recently been successfully validated with the fully equipped cryostat system~\cite{FirstLightKrypton2018}.

Tritium is injected at the centre of the beam tube with a pressure of \SI{3.4e-3}{\milli\bar} through \num{415} small orifices (each \SI{2}{\milli\metre} in diameter, see fig.~\ref{fig:injectionchamber}), resulting in an overall column density of tritium molecules $\mathcal{N}$ of \SI{5e21}{\per\metre\squared}. A stable inlet pressure is provided using a temperature and pressure stabilised buffer vessel at the beginning of the tritium feed line (see~\cite{Priester2015} for details). 

To reach the required gas flow retention in the spectrometer direction, two further pumping sections are attached: the DPS2~\cite{KATRIN2004,PhDSturm2010} (differential pumping) and the CPS~\cite{Gil2010} (cryogenic pumping). The latter relies on cryosorption of tritium on a cold surface~\cite{Eichelhardt2009}. 
The  \textbeta-decay electrons can pass the pumping sections as they are guided through magnetically. If they have enough energy to overcome the spectrometer retarding voltage they contribute to the measured spectrum. 
From this spectrum the neutrino mass will be extracted by fitting a  model function with several free parameters, making an accurate modelling of the spectrum of \textbeta-decay electrons leaving the source and transport section indispensable.

\paragraph{Modelling of \textbeta-decay electron spectra} \label{subsec:electron_spectrum}
In the spectral modelling, all energy loss processes of \textbeta-decay electrons reaching the detector need to be considered. Together with the transmission characteristics of the spectrometer they can be accounted for using the concept of a response function $R(E, U, \theta, z)$~\cite{Kleesiek2014,Kleesiek:2018mel}. Thus, the signal rate $\dot{N}(U)$ for one of the 148 pixels of the detector at spectrometer retarding voltage $U$ can be described as
\begin{equation}
\label{eq:rate}
	\dot{N}(U) \propto \int\limits_{-L/2}^{+L/2} n(z) \int\limits_{qU}^{\infty}\frac{\textrm{d}\Gamma}{\textrm{d}E} \cdot R(E,U,z)\,\textrm{d}E \,\textrm{d}z
\end{equation}
where $\frac{\textrm{d}\Gamma}{\textrm{d}E}$ denotes the differential rate of \textbeta-decay electrons at the time of decay and $n(z)$ the gas number density distribution along the longitudinal source beam tube symmetry axis $z$ with origin $z=0$ at the centre of the source.

One of the most important energy loss mechanisms to be included in the response function is inelastic scattering of electrons by gas molecules in the source. The probabilities $P_i(z,\theta)$ for an electron to scatter $i$-times depend on its pitch angle relative to the magnetic field at creation $\theta$ and can be computed using~\cite{Aseev2000} as
\begin{equation}
\label{eq:scattering_prob} 
	P_i(z, \theta)=\frac{\left(\mathcal{N}_\textrm{eff}(z,\theta)\cdot\sigma\right)^i}{i!}\mathrm{e}^{-\mathcal{N}_\textrm{eff}(z,\theta)\cdot\sigma},
\end{equation}
with $\sigma$ denoting the total inelastic scattering cross section and 
\begin{equation}
\label{eq:effective_rhod}
	\mathcal{N}_\textrm{eff}(z,\theta)=\frac{1}{\cos\theta}\mathcal{N}(z)=\frac{1}{\cos\theta}\int\limits_{z'=z}^{+L/2}n(z')\,\text{d}z'
\end{equation}
denoting the effective partial column density that accounts for increasing path lengths due to non-zero electron emission angle $\theta$. In the narrow energy window in which \katrin~will scan the tritium spectrum, the energy dependence of $\sigma$ can be dropped.

With eqs.~\eqref{eq:rate} to~\eqref{eq:effective_rhod} we have the motivation for extensive simulations of the gas dynamics in the source: we need simulations to compute the effective column density at different $z$-coordinates to correctly model the response function and thereby the count rate as measured by the detector. The importance is stressed by the fact that only the integral quantity $\mathcal{N}\cdot\sigma$ is accessible by measurement. By shooting electrons from an electron gun located in the rear section (see fig.~\ref{fig:katrin_overview}) through the source and measuring the electrons reaching the detector without scattering we can precisely ($\sim0.1\%$) determine $\mathcal{N}\cdot\sigma$~\cite{Babutzka2012}. In the following we present the gas dynamics calculations of the individual source components forming the gas model of the source section.
\section{Modelling of gas flow in the components of the source section}\label{sec:gasdynamics-theory-and-calculations}
The transport of a gas can be described using the kinetic Boltzmann equation, which in the absence of external forces can be written as~\cite{Sharipov2016}
\begin{equation}
\label{eq:BE}
\frac{\partial f}{\partial t}+ \vec{v}\cdot \nabla_r f = Q(f, \vec{v}).
\end{equation}
By inserting  the collision integral $Q(f,\vec{v})$ which accounts for binary intermolecular collisions, the Boltzmann equation can be solved for the velocity distribution function $f(t,\vec{r},\vec{v})$ which depends on time $t$, position $\vec{r}$ and velocity $\vec{v}$. The moments of $f$ are macroscopic variables such as density, temperature or bulk velocity.

A direct numerical solution of eq.~\eqref{eq:BE} for general conditions requires great computational effort so that some simplifications to the collision integral $Q$ are needed, defined by the underlying model equations. Model equations used in the present paper are the BGK equation, proposed by Bhatnagar, Gross and Krook~\cite{Bhatnagar1954} (isothermal continuum flow), and the S-model by Sharkov~\cite{Shakhov1968} (non-isothermal continuum flow). 

Besides the numeric solution of eq.~\eqref{eq:BE}, there are approaches based on Monte Carlo calculations such as the test particle Monte Carlo (TPMC) method~\cite{Davis1960,Bird1963} (only particle-wall interaction, suited for molecular flow) and the direct simulation Monte Carlo (DSMC) method~\cite{Bird1963} (particle-particle and particle-wall interaction, suited for transitional flow). In order to cover the complete range of pressure regimes present in the \katrin~source section, those two Monte Carlo methods are used complementary to the model equations.

\begin{figure}
	\includegraphics[width=1\textwidth]{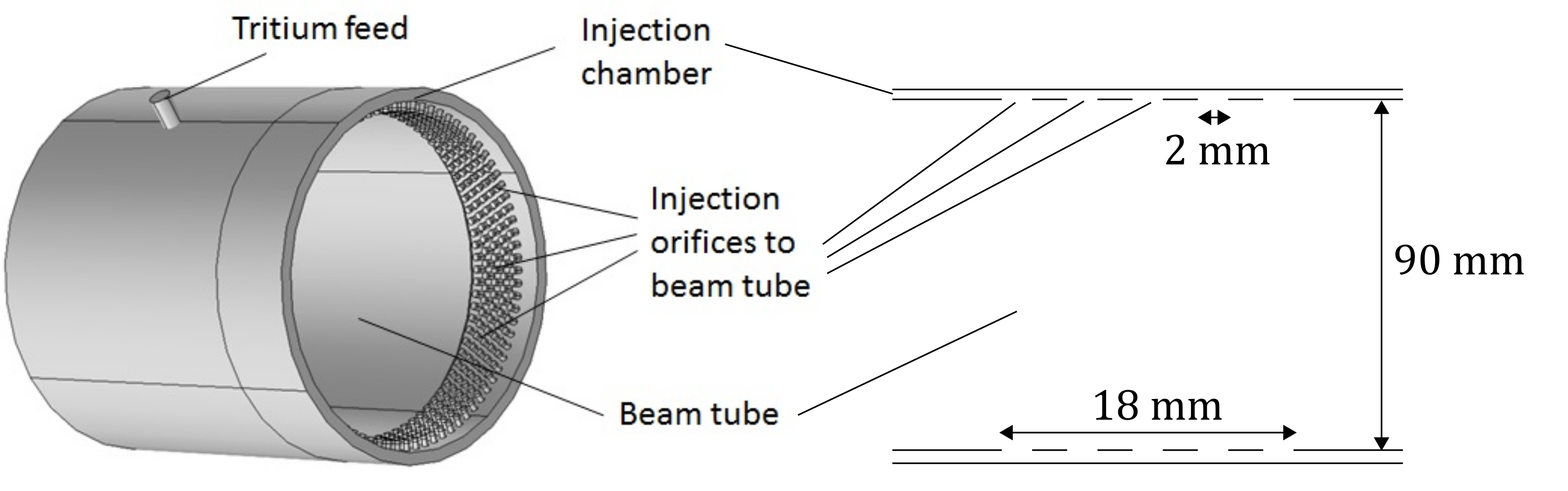}
	\caption{Injection chamber surrounding the WGTS beam tube with the feed capillary and the 415 injection orifices. A sketch of the longitudinal cross section is shown on the right (not to scale). Here we can see that the injection chamber is a space between two concentric cylinders.}
	\label{fig:injectionchamber}
\end{figure}

Further simplifications of eq.~\eqref{eq:BE} can be applied depending on the gas flow regime, which can be classified in terms of the rarefaction parameter $\delta$. This is inversely proportional to the equivalent free path $\lambda$  
\begin{equation}
\label{eq:delta}
\delta=\frac{a}{\lambda} \qquad \text{, }\lambda=\frac{\eta v_{\textrm{m}}}{p}, 
\end{equation}
with characteristic dimension $a$, most probable speed $v_\textrm{m}$, pressure $p$ and viscosity $\eta$.
Typically,  three regimes can be distinguished:
\begin{itemize}
	\item $\delta \gg 1$: hydrodynamic or continuum regime; gas flow can be described by continuum mechanics.
		\item $\delta \approx 1$: transitional regime; continuum mechanics is not valid and intermolecular collisions are not negligible.
	\item $\delta \ll 1$: free molecular regime; intermolecular collisions can be neglected since $\lambda \gg a$.
\end{itemize}
\begin{figure}
	\includegraphics[width=1.0\textwidth]{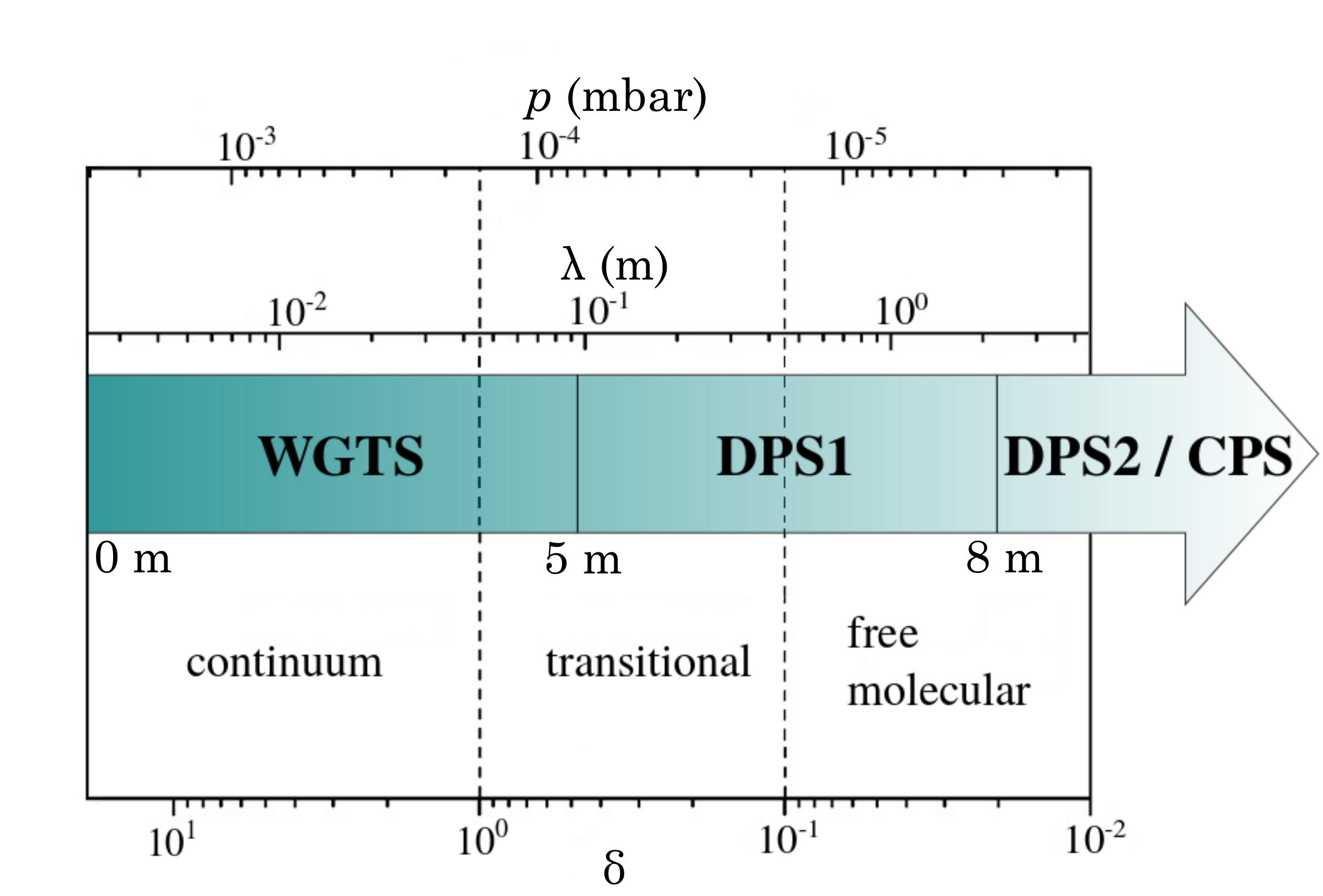}
	\caption{Range of rarefaction parameter $\delta$  with associated rarefaction regimes, corresponding pressure $p$ and equivalent mean free path $\lambda$, values for \katrin~measurement conditions. A constant tube radius of \SI{45}{\milli\metre} in the source and transport section is assumed for illustration purposes.
		\label{fig:pressure_regimes}}
\end{figure}
The source section of \katrin~covers the whole range of rarefied gas flow regimes (compare fig.~\ref{fig:pressure_regimes}).
Tracing the trajectory of a \tritiummol~molecule towards the spectrometer section, it starts in the hydrodynamic regime at the inlet chamber in the middle of the source (see fig.~\ref{fig:injectionchamber}), enters the transitional regime while still in the WGTS beam tube  and reaches free molecular flow in the second  pump port of the  DPS1. Because of the combination of widely disparate rarefaction regimes, different approaches need to be used to describe the particular components of the source  section.

\begin{figure}
  \includegraphics[width=1.0\textwidth]{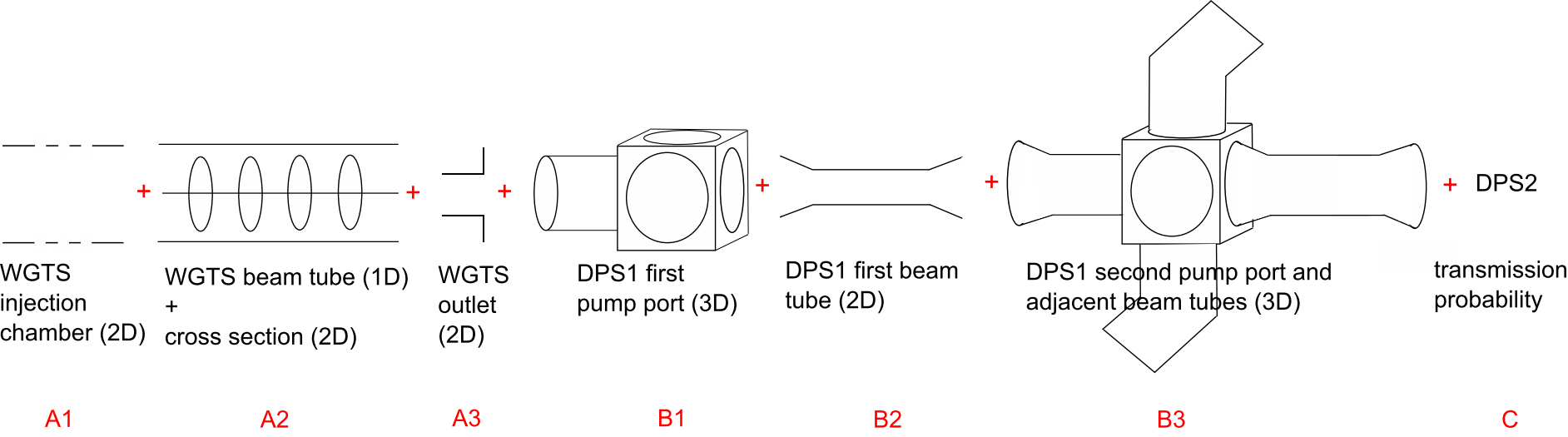}
  \caption{Schematic overview of the geometric model of WGTS and DPS sections used for the gas flow simulation (not to scale). The corresponding dimensionality of the gas dynamics model in the respective domains is also indicated.}
  \label{fig:scheme_gas}
\end{figure}

Moreover, splitting the calculation of gas flow in the source section is  motivated by the complex geometries of the pump ports, which are quite demanding in terms of computational resources.
Some domains are simplified to two- or even one-dimensional geometric representations as summarised in fig.~\ref{fig:scheme_gas}.

A calculation of gas flow through the \katrin~source and transport section has been presented by Malyshev et al.~\cite{Malyshev2009}. However, important effects related to tritium injection and outflow as well as to temperature anisotropies were not considered previously. Furthermore, the TPMC method, suitable for the description of molecular flow, was applied by Malyshev et al. in the first pump port and beam tube sections of the DPS1 where the gas flow is still transitional. Moreover, the DPS2 has undergone significant design modifications with respect to the model used by Malyshev et al.~\cite{Malyshev2009}. Thus, the calculation needs to be refined and adapted for the new design.

In the following, the refined gas dynamics model of the source section  is presented. The investigations of local gas flow disturbances such as injection and outlet geometry as well as anisotropic temperature gradients are illustrated in detail to show their effect on the column density and to validate important simplifications. 

\subsection{Density distribution in the WGTS beam tube (A1-A3)}\label{sec:Density_WGTS} 
About 99\% of the  total column density is situated in the central \SI{10}{\metre} WGTS beam tube. Therefore, the gas flow through this domain needs to be calculated accurately. Here temperature non-uniformities as well as inlet and outlet effects need to be considered more thoroughly than in the simulation of the other parts of the source system. Because of the large length-to-radius ratio of about \num{100}, a one-dimensional fully developed flow (no disturbances by inlet or outlet) approach is suitable for the main part of the tube (region A2 in fig.~\ref{fig:scheme_gas}) to reduce complexity. Inlet and outlet regions are modelled separately in two and three dimensions (regions A1, A3 and B1 in fig.~\ref{fig:scheme_gas}), as their flow profiles deviate from the fully developed case. 
 
For the 1D main beam tube calculation along the $z$ axis the method described in ref.~\cite{Sharipov1997,Sharipov1998} is used. The mass flow rate $\dot{M}$ can be represented as
\begin{equation}
\label{eq:mass_flow_globalG} 
\dot{M}=\frac{\pi r_0^3}{v_\textrm{m}}\left(-G_\text{P}(\delta)\frac{\textrm{d}p}{\textrm{d}z} + G_\text{T}(\delta)\frac{p(z)}{T(z)}\frac{\textrm{d}T}{\textrm{d}z}\right),
\end{equation}
where $r_0$ is the beam tube radius, $p(z)$ is the local pressure, $T(z)$ is the local temperature and $\delta$ is the rarefaction parameter defined by eq.~\eqref{eq:delta} with $r_0$ as the characteristic dimension. The Poiseuille $G_\text{P}$ and thermal creep $G_\text{T}$ coefficients are functions of the rarefaction parameter~\cite{Sharipov2016}:
\begin{equation}
	G_\text{P} = \frac{8}{3\sqrt{\pi}}\frac{1+0.04\,\delta^{0.7}\ln\delta}{1+0.78\,\delta^{0.8}}+\left(\frac{\delta}{4}+1.018\right)\frac{\delta}{1+\delta}
	\label{eq:GP}
\end{equation}
\begin{equation}
	G_\text{T} = \left\{
					\begin{array}{l l l}
						&\frac{4}{3\sqrt{\pi}}+0.825(1+\ln\delta)\delta-(1.18-0.61\ln\delta)\delta^2  & \text{for}\ \delta\leq1\text{,}\\ 
						&\frac{1.175}{\delta}-\frac{1.75}{\delta^2}+\frac{1.47}{\delta^3}-\frac{0.5}{\delta^4}& \text{for}\ \delta>1\text{.} 
					\end{array}
				\right.
\end{equation}
Moreover, $G_\text{P}$ and $G_\text{T}$ are determined by the gas-surface interaction which is taken into account via the accommodation coefficient $\upalpha$ describing the gas-surface interaction. Since $\upalpha$ only weakly affects the column density, it is appropriate to assume full accommodation which is $\upalpha=1$. 

We will make use of both non-isothermal flow and isothermal flow:
\begin{itemize}
	\item Isothermal flow: the Poiseuille coefficient is integrated with respect to $z$ from the injection cross section ($z=0$) to outlet cross section ($z=L/2$)  so that the mass flow rate reads
	\begin{equation}
	\label{eq:mass_flow_globalG_P} 
	\dot{M}=\frac{\pi r_0^3 p_\textrm{in}}{v_m\delta_\textrm{in}\left(L/2\right)}\int_{\delta_\textrm{out}}^{\delta_\textrm{in}}G_\text{P}(\delta)\,\textrm{d}\delta,
	\end{equation}
	where $L$ is the beam tube length,  $p_\textrm{in}$ is the injection pressure and $\delta_\textrm{in}$ and $\delta_\textrm{out}$ are rarefaction parameters in the injection and outlet sections, respectively.
	
	\item Non-isothermal flow: the distribution $\delta(z)$ can be calculated from eq.~\eqref{eq:mass_flow_globalG} by applying a finite difference scheme as used in ref.~\cite{Sharipov1996,Sharipov1997,Sharipov1998}.
\end{itemize}
In order to convert the rarefaction parameter distribution into a pressure distribution, the viscosity of the tritium gas needs to be known. It is derived from hydrogen and deuterium using the mass ratio of the isotopologues. Discrepancies occur due to quantum effects at low temperature. Comparing hydrogen~\cite{Assael1986H2} and deuterium~\cite{Assael1987D2} viscosities from measurement to the approximation formula 
\begin{equation}
\label{eq:deuterium_viscosity}
\upeta_\textrm{D$_2$}=\sqrt{\frac{m_\textrm{D$_{2}$}}{m_\textrm{H$_2$}}}\upeta_\textrm{H$_2$}
\end{equation}
showed that eq.~\eqref{eq:deuterium_viscosity} provides a 7\% overstated value of $\upeta_{\text{D}_2}$. We therefore assume that eq.~\eqref{eq:deuterium_viscosity} applied to T$_2$ provides a 5\% overstated value of $\upeta_{\text{T}_2}$, with an uncertainty of 7\%. Thus the tritium viscosity at \SI{30}{\kelvin} is approximated by~\cite{Hoetzel2012}
\begin{equation}
\label{eq:tritium_viscosity}
	\upeta_\textrm{T$_2$}=\num{0.95}\cdot\sqrt{\frac{m_\textrm{T$_{2}$}}{m_\textrm{D$_2$}}}\upeta_\textrm{D$_2$}\approx\SI{2.425e-6}{\pascal\second}.
\end{equation}
Due to a small longitudinal asymmetry (about \SI{7}{\centi\metre} difference in length) of the WGTS beam tube, calculations need to be done for both flow directions separately (each starting from the centre of the inlet A1 of fig.~\ref{fig:scheme_gas}). 

\begin{figure}
	\subfigure[Longitudinal velocity profile (1D)]{
		\centering
		\includegraphics[width=0.48\textwidth]{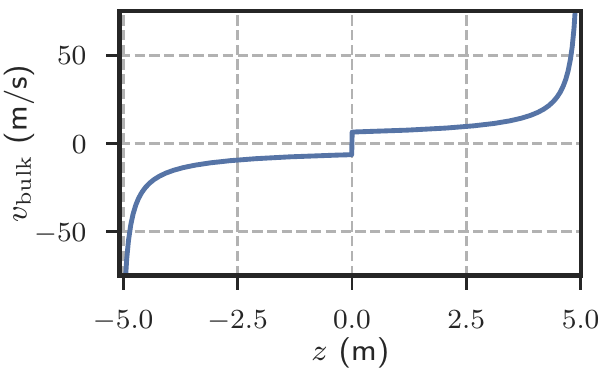}
		\label{fig:longitudinal_velocity}
	}
	%\quad
	\subfigure[Radial velocity profile (2D)]{
		\centering
		\includegraphics[width=0.48\textwidth]{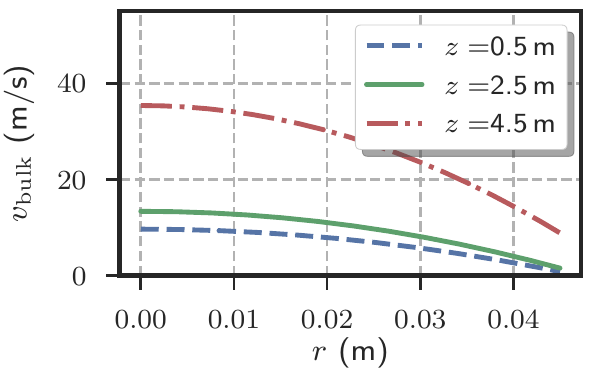}
		\label{fig:radial_velocity}
	}
	\caption[Longitudinal and radial velocity profile]{Longitudinal~\ref{fig:longitudinal_velocity} and radial~\ref{fig:radial_velocity} velocity profile. Gas streaming towards the detector has positive bulk velocity while gas streaming to the rear side has negative bulk velocity.}
	\label{fig:velocity}
\end{figure}

The direction of flow can also be visualised by the longitudinal velocity profile (radially averaged) in fig.~\ref{fig:longitudinal_velocity}, which shows negative bulk velocities for gas going to the rear side and positive for gas going to the detector side. Limitations to the 1D calculation are shown by the 2D cross sections: Gas streaming on the $z$-axis has higher bulk velocity than gas streaming close to the tube walls, which can be seen from fig.~\ref{fig:radial_velocity}.

\begin{figure}
   \subfigure[Longitudinal density deviation]{\includegraphics[width=0.49\textwidth]{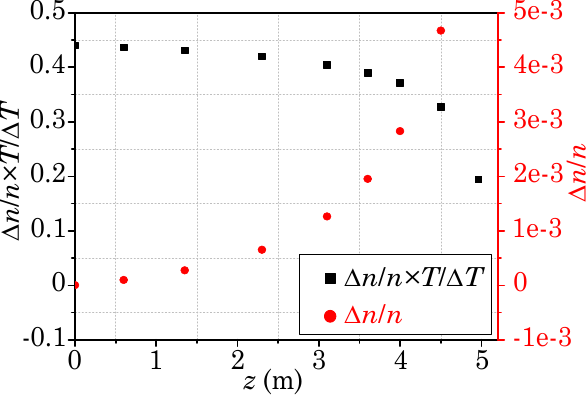}\label{fig:3D_WGTS_flow_b}}\quad
    \subfigure[Azimuthal density deviation]{\includegraphics[width=0.49\textwidth]{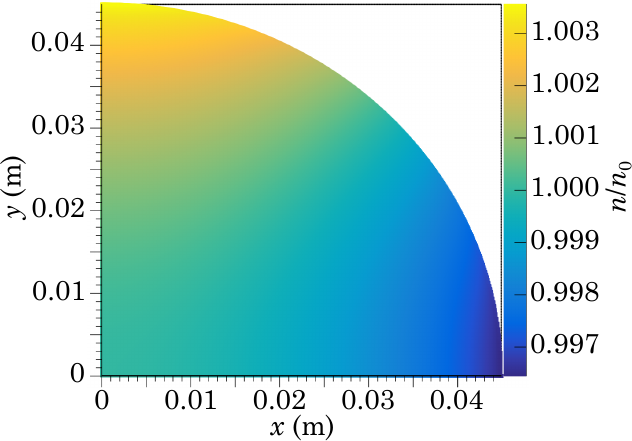}\label{fig:3D_WGTS_flow_a}}
    \caption{Calculation of non-isothermal WGTS flow using the temperature distribution from eq. (\ref{eq:dT_phi}).  In a) the maximal change in relative density for different positions at the $z$ axis is depicted for a fixed temperature deviation $\Delta T=\SI{1}{\kelvin}$ (in red) and scaled with the temperature distribution  (in black). Note  the corresponding different axes. In b) the relative density of one quadrant of a cross section at $\SI{3.6}{\metre}$ distance to the centre of the WGTS, normalised to the one-dimensional calculation $n_0$, is shown.}
\end{figure}

Another limitation of the 1D calculation is azimuthal temperature variation since this causes radial and azimuthal flow and thus changes the density profile: The parts of the walls not in contact with the beam tube cooling pipes can get warmed to a small extent. 
The heat flux mainly enters through thermal radiation in the pump ports at both ends of the WGTS beam tube~\cite{Wolf2012}. 
Due to a  special cooling concept and thermal shielding, the magnitude of longitudinal and azimuthal temperature gradients is limited to about \SI{1}{\kelvin}~\cite{Grohmann2009,Grohmann2013}. 
Based on the \textit{Demonstrator} measurements described in ref.~\cite{Grohmann2013} the form of the beam tube temperature profile $T(\phi,z)$, with $\phi$ denoting the azimuthal angle, can be approximated as
\begin{equation}
 \label{eq:dT_phi}
 T(\phi,z)=T(z)+\Delta T(z)\sin^2(\phi).
\end{equation} 
Assuming small pressure and temperature gradients, longitudinal and azimuthal flow can be handled separately~\cite{Sharipov2009circ}.
The resulting relative density deviation is depicted in fig.~\ref{fig:3D_WGTS_flow_b}. Assuming a maximal temperature difference $\Delta T= \SI{1}{\kelvin}$, the average column density difference, compared to the 1D isothermal calculation, is \num{0.15}\%. 
Using the  precalculated cross sectional flow distributions, an example of which is shown in fig.~\ref{fig:3D_WGTS_flow_a}, it is possible to correct for a given temperature profile.

Further deviations from the one-dimensional fully developed tube flow occur due to end-effects in inlet and outlet regions.
Thus, two-dimensional models with a length of \SI{40}{\centi\metre} for the inlet region (A1 in fig.~\ref{fig:scheme_gas}; a sketch of the model geometry is depicted in fig.~\ref{fig:injectionchamber}) and \SI{20}{\centi\metre} for the outlet region (A3 in fig.~\ref{fig:scheme_gas}) are built. The gas flow in these regions is modelled with the BGK model equation in its linearised form. 

\begin{figure}
    \includegraphics[width=1\textwidth]{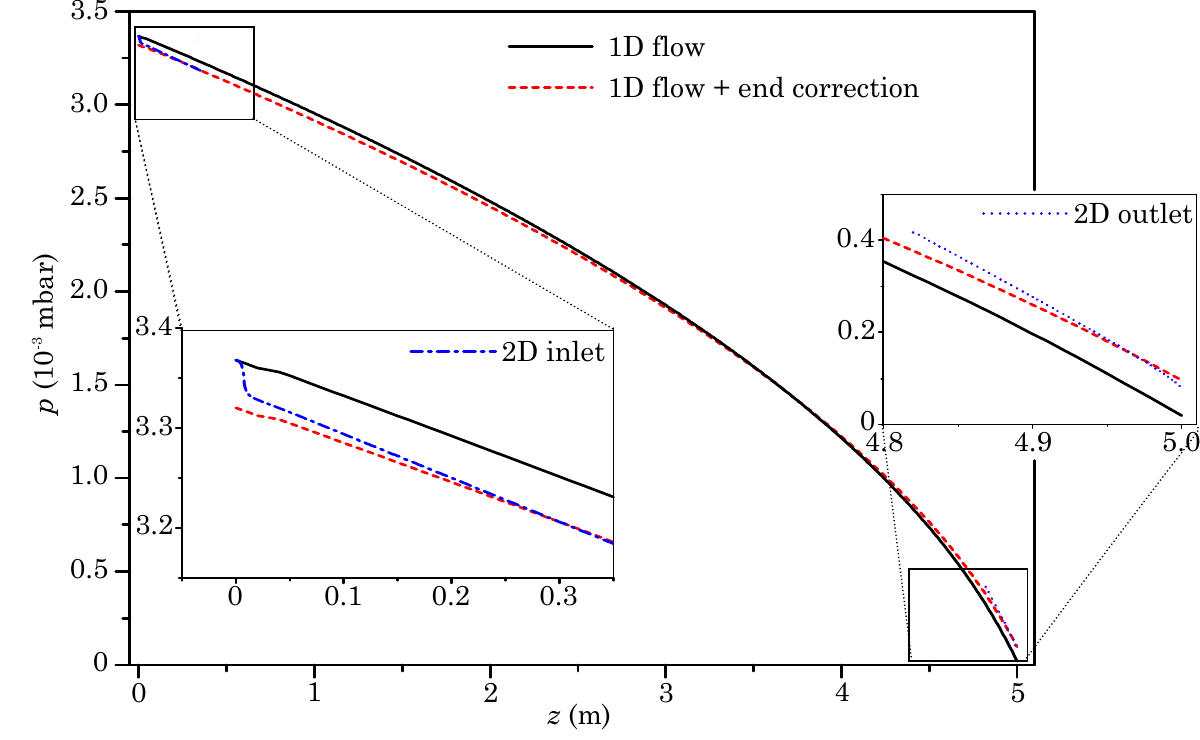}
     \caption{Pressure distribution along the WGTS beam tube axis. The one-dimensional calculation without  end-effects is plotted in black. The end-effect corrected distribution is plotted in red. Radial averaged distributions from  two-dimensional inlet and outlet model are plotted in blue and shown in the inset of inlet and outlet region.}
     \label{fig:end_effect1D}
\end{figure}

For the simplified 2D inlet calculations (A1 in fig.~\ref{fig:scheme_gas}), the pressure gradient in radial direction reaches 0 about \SI{25}{\milli\metre} after injection~\cite{Kuckert2016}, which also means we can only observe local distortions in fig.~\ref{fig:end_effect1D}; the same holds for the 2D outlet.
1D longitudinal density distributions for both models are depicted in fig.~\ref{fig:end_effect1D} along with the beam tube calculation results.

To be used in the \textbeta-spectrum modelling, inlet pressure and temperature conditions need to be variable according to experimental conditions. Since the influence of the inlet and outlet regions is only local, so-called end-effect corrections of the mass flow rate and the Poiseuille coefficient can be calculated according to the method described in refs.~\cite{Sharipov1998,Pantazis2013end,Pantazis2014end}.
It is based on a correction of the tube length $\Delta L$ that is obtained from the 1D and 2D calculation results.
Now the mass flow rate from eq.~\eqref{eq:mass_flow_globalG_P} can be modified using
the end corrections in the injection and outlet  sections $\Delta L_\textrm{in}$ and $\Delta L_\textrm{out}$, respectively:
\begin{equation}
 \label{eq:Poisseulle_corrected}
 \dot{M}\propto\frac{L}{L+\Delta L_{\textrm{in}}+\Delta L_{\textrm{out}}\cdot\left(\delta_{\textrm{in}}-\delta_{\textrm{out}}\right)} \int_{\delta_{\textrm{out}}}^{\delta_{\textrm{in}}}G_\text{P}(\delta)\,\textrm{d}\delta.
\end{equation}
A deviation of about \num{5}\%  for mass flow rate and throughput is obtained comparing  end-corrected and uncorrected one-dimensional results. Nevertheless, the overall column density deviation is smaller than \num{1}\%, since inlet and outlet effect cause  opposite density changes that partially cancel each other, as can be seen in fig.~\ref{fig:end_effect1D}.

\subsection{Density distribution in the DPS1 first pump port (B1)}
\label{subsec:PP1}
\begin{figure}
   \subfigure[] {\includegraphics[width=0.49\textwidth]{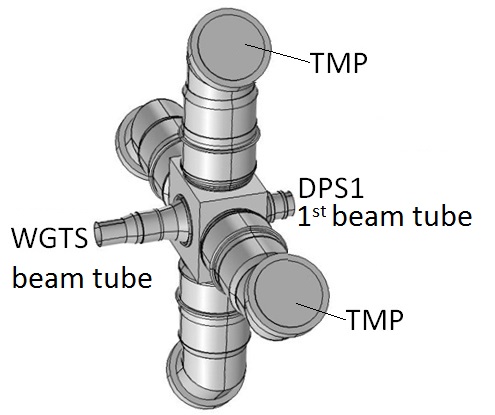}\label{fig:1stPumpPortModel_a}}\quad
   \subfigure[]{\includegraphics[width=0.49\textwidth]{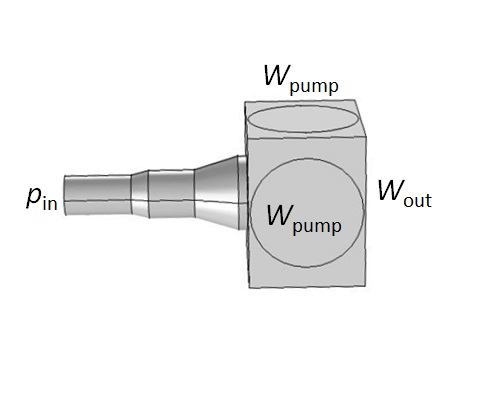}\label{fig:1stPumpPortModel_b}}
    \caption{Three-dimensional geometry of the first pump port of the DPS1. a) shows the complicated geometry. b) shows the simplified geometry used for the pump port gas flow simulation. Here the tubes are represented by transmission probabilities $W_{\textrm{pump}}$ and $W_{\textrm{out}}$.}
\end{figure}

The density distribution in the first pump port of the DPS1 (B1 in figure~\ref{fig:scheme_gas}) needs to be computed accurately to determine the relative outlet pressure of the WGTS beam tube, an input parameter for the WGTS beam tube density calculation described above, and to calculate the gas flow reduction factor. A three-dimensional model is required to investigate the gas flow through the complex geometry depicted in fig.~\ref{fig:1stPumpPortModel_a}. The rarefaction at the beginning of the first pump port is $\delta\approx\num{0.5}$. Since the gas is still in the transition regime the TPMC method as used by Malyshev et al.~\cite{Malyshev2009} is not suitable. A DSMC approach with \num{e7} model particles is chosen, as the model cannot be calculated analytically. The solution procedure is further described in ref.~\cite{Sharipov2004numerical} and~\cite{Varoutis2008}.

Despite the high temperature (about \SI{330}{\kelvin}) at the rotor blades of the pumps, the pump port itself is expected to have an almost homogeneous temperature of about \SI{30}{\kelvin} which allows an isothermal approach~\cite{Kuckert2016}.  
In order to match the three-dimensional pump port simulation to the one-dimensional beam tube calculation, both geometries have an overlapping region of \SI{0.32}{\metre} length.
The tubes that are connected to the pump port are represented by semitransparent boundaries to simplify the geometry. The transmission probabilities $W$ of these boundaries can be approximated using their length-to-radius ratio~\cite{Berman1965,Lund1966flow}. This results in $W_{\textrm{out}}\approx\num{0.1}$ for the tube connecting the first and second pump port, and $W_{\textrm{pump}}^{\textrm{B1}}\approx\num{0.36}$ for the tube connecting the pump port to the TMP.

\begin{figure}
	\includegraphics[width=1.\textwidth]{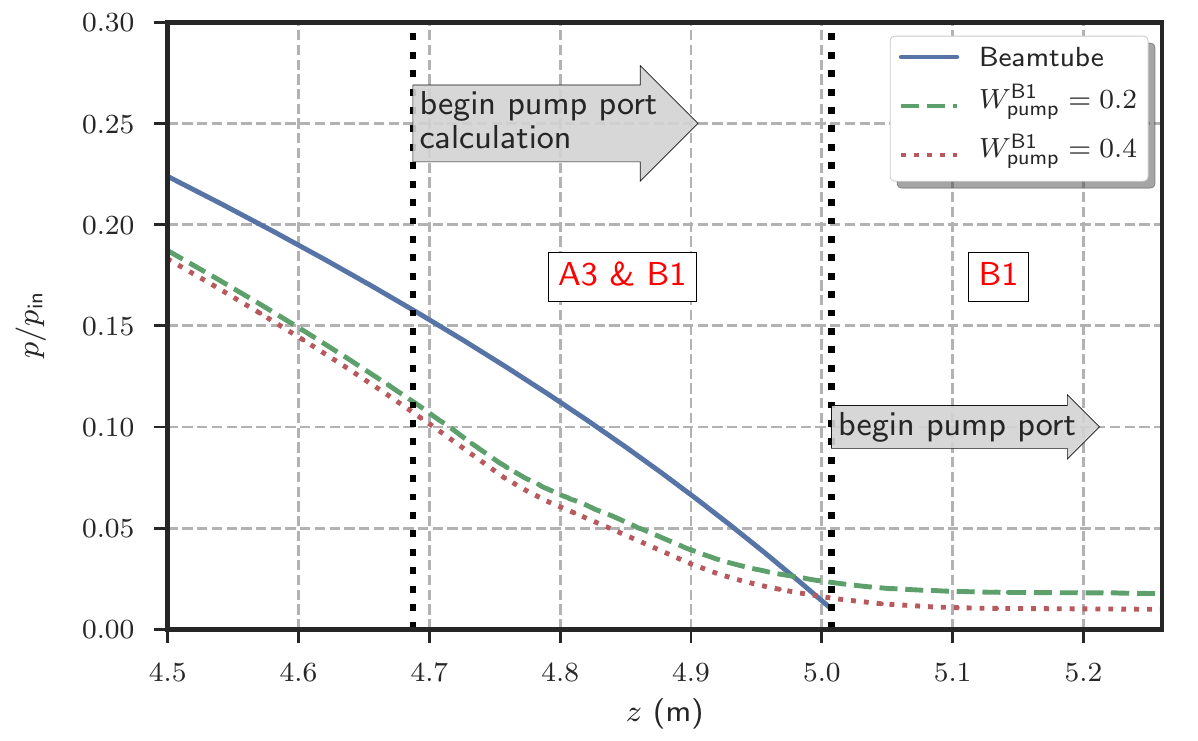}
	\caption{Relative pressure distribution for the three-dimensional pump port simulation with two values of $ W_\text{pump}^\text{B1}$. The pressure distribution from the one-dimensional beam tube only calculation is plotted in solid blue. The smooth transition from 1D to 3D is marked by \textit{begin pump port calculation}}
 	\label{fig:1st_pp_simulation}
\end{figure}

Since the latter tube is bent and has a conical shape, its  transmission probability will likely be reduced so calculations are carried out for two extremes: $W_{\textrm{pump}}^{\textrm{B1}}=\num{0.4}$ and \num{0.2}, see fig.~\ref{fig:1st_pp_simulation}. 
The density distribution and column density inside the first pump port are affected significantly, as can  be seen in fig.~\ref{fig:1st_pp_simulation}, while the effect on the total column density is small (below \num{0.04}\%), since the  pump port density contribution is only about \num{0.1}\%.

The calculated density at the end of the WGTS beam tube is about \num{2}\% of the inlet density and the flow-reduction factor of the first pump port (for $W_{\textrm{pump}}^{\textrm{B1}}=\num{0.2}$) is about \num{33}.  

\subsection{Density distribution in the DPS1 first beam tube (B2)}
\label{subsec:BT1}
Aiming for a column density uncertainty of \num{0.2}\% the outer source region still needs to be considered in eq.~\eqref{eq:scattering_prob}, despite its relatively low tritium amount.  
The flow in the first tube of DPS1, which has a length-to-radius ratio of about 20 and  connects the first and second pump ports, is computed using a two-dimensional geometry (B2 in fig.~\ref{fig:scheme_gas}) to include end-effects by adding a pump port model at each end. A transitional flow approach has to be used, since the rarefaction parameter is between \num{0.2} and \num{0.4}. This domain is simulated using the transitional flow interface of COMSOL Multiphysics (version 5.0) \cite{COMSOL}. The BGK model equation is applied and solved by adopting the discrete velocity method \cite{Carleman1957,Broadwell1964}. 

To evaluate the validity of the isothermal flow assumption required for the BGK model equation~\cite{Sharipov1998}, the fraction of non-isothermal temperature-driven flow is approximated with the help of eq.~\eqref{eq:mass_flow_globalG}, assuming a conservative temperature difference of \SI{6}{\kelvin}~\cite{Kuckert2016}. The resulting relative flow-rate difference and discrepancy in pressure between isothermal and non-isothermal flow are about \num{5}\%. Since this domain's column density  contributes less than \num{0.3}\% to the total value, the corresponding column density modelling uncertainty is smaller than \num{1e-4} and the effect of the non-isothermal flow is neglected for the overall column density modelling uncertainty budget.

\subsection{Density distribution in the DPS1 second pump port and adjacent beam tubes (B3)}
\label{subsec:PP2}
For the last part of the source geometry, including the second pump port of the DPS1 and the adjacent  tube to the DPS2, a molecular flow approach can be used ($\delta<\num{0.1}$) along with a more resource-intensive 3D model. To account for the molecular beaming effect, the  model contains the previously mentioned tube entering the second pump port. The temperature changes significantly from about \SI{30}{\kelvin} at the beginning of the domain to room temperature at its end, thus making an isothermal TPMC approach unsuitable. 

Instead, the molecular flow interface of the COMSOL microfluidics module~\cite{COMSOL} is applied here. It makes use of the angular coefficient method~\cite{Kersevan2009} which allows the explicit inclusion of temperature differences.
As extension to the model used in sec.~\ref{subsec:PP1}, the new model contains the pumping ducts. 

The pump itself is replaced by a partially absorbent surface with a gas-temperature-dependent transmission probability $W_{\textrm{pump}}^{\textrm{B3}}$ of \num{0.3} as evaluated at \SI{275}{\kelvin}.
The transmission probability at the outlet boundary to the DPS2 is set to \num{0.2} according to the calculations in~\cite{Jansen2015}.
 
As a result, density and gas flow reduction  factors of \num{4.7} and \num{11.7} are computed and the column density of the modelled domain accounts for \num{0.03}\% of the total column density.

\subsection{Complete gas model: combining the different domains}
\label{subsec:gascomplete}
\begin{figure}[ht]
	\includegraphics[width=1.\textwidth]{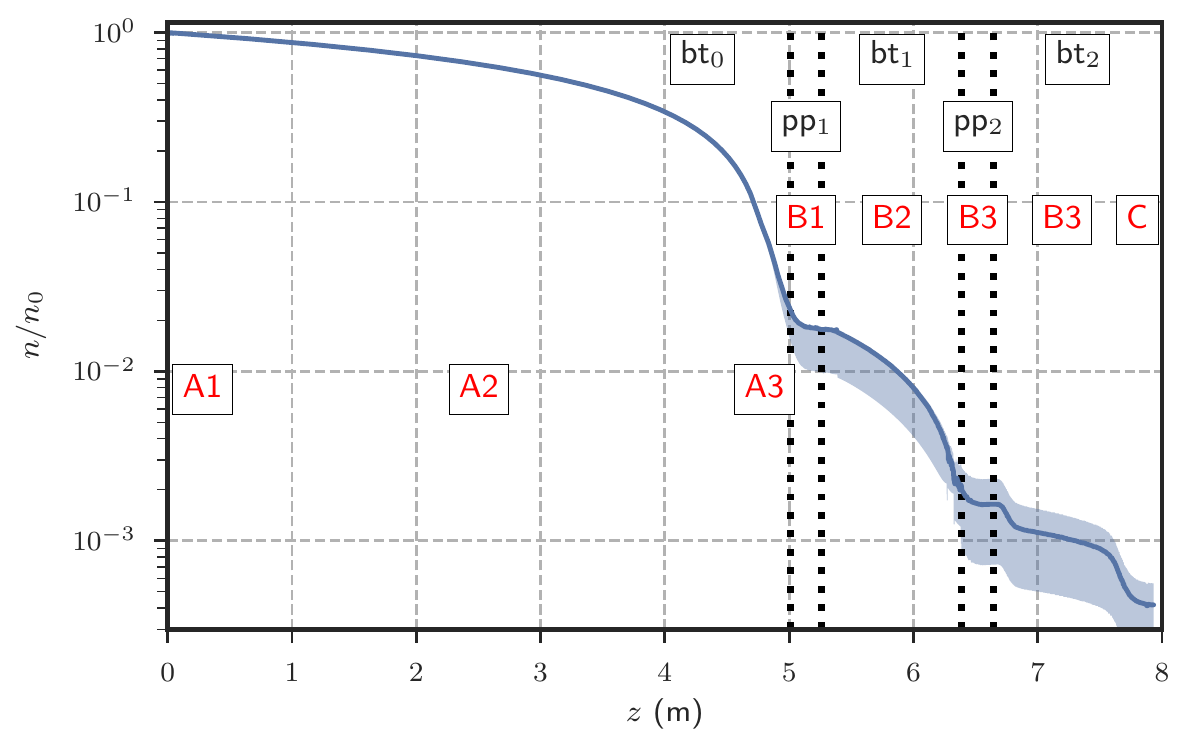}
	\caption{Relative density distribution in the source section along the beam tube axis in the front direction, based on calculations for the WGTS beam tube, the two DPS1-F pump ports, and connecting tubes. The uncertainties of the density distributions in these domains (described in the text) are assumed to propagate to the adjacent components in the direction of flow (marked by the shaded area). The error band becomes asymmetric because the transmission probability $W_\textrm{pump}^{\textrm{B1}}$ to the TMP in the first pump port might be larger (results are given for $W_\textrm{pump}^{\textrm{B1}}=\num{0.2}$) which would result in a larger density reduction. 
 	\label{fig:full_WGTS_pressure} }
\end{figure}

To form a complete  gas model of the source section all domain calculations need to be connected. The inlet density is defined solely in the calculation of the gas flow in the main beam tube and the density profiles for all subsequent domains are scaled accordingly. The overall model column density can therefore be adjusted to the measured value. 
This composite density distribution for the complete source section is depicted in fig.~\ref{fig:full_WGTS_pressure}.  

Overall reduction factors for density and gas flow are about \num{2000} and \num{400}, respectively. The reduction factors for the particular domains in the rear and front directions are summarised in tab.~\ref{tab:flow_results}.  
Including the DPS2 reduction factor of \num{4.5e5}, as calculated in~\cite{Jansen2015}, we obtain a gas flow reduction factor of \num{1.7e8} for the  combined differential pumping sections in the forward direction.
 
The presented gas model allows us to calculate  the density distribution in the whole  \katrin~source section. Though ranging from the continuum to the free molecular regimes, it is adjustable with respect to inlet pressure and tube wall temperature. The uncertainty of the column density calculation is governed by the modelling of gas flow in the central WGTS beam tube; contributions from subsequent domains were shown to be one or two orders of magnitude smaller. 
The following section describes how the column density will be determined and monitored during \katrin~measurements based on the gas model.
\begin{table}[htbp]
  \centering
  \caption{Density and flow reduction factors, $x_n$ and $x_q$, for all simulated domains of the source section, also stating the proportion $\mathcal{N}$ of total column density $\mathcal{N}_0$ per domain. Differences in front and rear distributions are due to a small longitudinal WGTS beam tube asymmetry. Values for the first pump port are given for $W_{\textrm{pump}}^{\textrm{B1}}=\num{0.2}$. The abbreviations `bt' and `pp' represent beam tube and pump port sections, respectively.}
    \begin{tabular}{p{4.5cm}cccc}
    \toprule
      Domain    & $n_{\textrm{out}}/n_{\textrm{in,WGTS}}$ & $\mathcal{N}/\mathcal{N}_{\textrm{0}}$ & $x_n$ & $x_q$ \\
    \midrule
    DPS1-R 2\textsuperscript{nd} pp+bt (B3) & \num{5.2e-4} & \num{3e-4} & \num{4.7} & \num{11.7} \\
    DPS1-R 1\textsuperscript{st} bt (B2) & \num{2.23e-3} & \num{2.6e-3} & \num{11.1} & - \\
    DPS1-R 1\textsuperscript{st} pp (B1) & \num{0.025} & \num{1.1e-3} & \num{1.3} & \num{33} \\
    WGTS bt rear (A1-A3) & \num{0.033} & \num{0.499} & \num{30.4} & - \\
    WGTS bt front (A1-A3) & \num{0.034} & \num{0.492} & \num{29.2} & - \\
    DPS1-F 1\textsuperscript{st} pp  (B1) & \num{0.026} & \num{1.1e-3} & \num{1.3} & \num{33} \\
    DPS1-F 1\textsuperscript{st} bt (B2) & \num{2.34e-3} & \num{2.7e-3} & \num{11.1} & - \\
    DPS1-F 2\textsuperscript{nd} pp+bt (B3) & \num{5e-4} & \num{3.1e-4} & \num{4.7} & \num{11.7} \\
    \midrule
    Front direction source section & \num{5e-4} & \num{0.496} & \num{1980} & \num{386} \\
    \bottomrule
    \end{tabular}%
  \label{tab:flow_results}%
\end{table}

\section{Column density and its role in the neutrino mass measurement}\label{sec:source-modeling}
The spectral distribution of electrons leaving the source is influenced significantly by the column density that controls the scattering probabilities, as discussed in section~\ref{sec:wgts}: The parameter of interest is $\mathcal{N}\cdot\sigma$, which needs regular experimental monitoring to ensure uncertainty below 0.2\%~\cite{KATRIN2004}. In the following, two different procedures will be presented: 1) the determination of the absolute value of $\mathcal{N} \cdot\sigma$ and 2) the measurement of $\mathcal{N}$ fluctuations induced by changes of operating parameters.

\subsection{Absolute value determination}\label{subsec:egun_measurement}
The absolute value of $\mathcal{N}\cdot\sigma$~can in general be determined either based on gas flow simulations as presented in section~\ref{sec:gasdynamics-theory-and-calculations} or by a dedicated measurement. 
The simulation-based value is obtained by multiplication of the calculated longitudinally integrated density profile $\mathcal{N}$~with the literature value of the total scattering cross section $\sigma=\SI{3.40(7)e-18}{\centi\metre\squared}$~\cite{Aseev2000} for electron energies at the tritium endpoint.

Since about \num{99}\% of the total column density is situated inside the WGTS beam tube (see tab.~\ref{tab:flow_results}), it is sufficient to consider the accuracy of the gas density there. The uncertainty of the model equation used for this domain exceeds the \katrin~requirement of 0.2\%: In refs.~\cite{Graur2011} and~\cite{Sharipov2009} differences not larger than \num{2.5}\% are derived for the calculated flow rate coefficients $G_\text{P}(\delta)$ and $G_\text{T}(\delta)$	by comparing results from different modelling techniques (BGK equation, S-Model and  DSMC method) in the transition regime at $\delta \approx \num{1}$. Moreover, in the hydrodynamic regime the different model solutions agree within the numerical uncertainty of 0.5\%. As most parts of the WGTS beam tube flow are in the hydrodynamic regime ($\approx\SI{8}{\metre}$ out of the central \SI{10}{\metre} WGTS beam tube), the average rate coefficient uncertainty can be taken to be smaller than \num{2}\%. Including the \num{2}\% on the literature value of the scattering cross section~\cite{Aseev2000} even would increase the uncertainty on $\mathcal{N}\cdot\sigma$ beyond 2\%. This accuracy would not match the requirement of \num{0.2}\%, making the calculation method unsuitable for the determination of the absolute value of $\mathcal{N}\cdot\sigma$.

Following a different approach, $\mathcal{N}\cdot\sigma$ is determined by measurement using an electron gun (e-gun, similar to the one in ref.~\cite{Behrens2017}) that is installed at the rear end of the \katrin~beam line. A mono-energetic beam of electrons with energy $E_{\textrm{e0}}$  is sent through the WGTS filled with gas (column density $\mathcal{N}$). 

To determine the initial rate of the beam $\dot{N}_{\textrm{e}}(0)$, it is also sent through an evacuated WGTS (pressure below \SI{1e-6}{\milli\bar}, so that the probability of the electrons to scatter on residual gas can be neglected) in a separate measurement.

The spectrometer is set on a retarding potential \SI{5}{\volt} below $E_\text{e0}$. Electrons that have scattered inelastically with gas molecules in the source lose at least \SI{9}{\electronvolt} in energy~\cite{KATRIN2004,Aseev2000}. This prevents them from overcoming the potential barrier and thus only unscattered electrons are detected. A comparison of the rate of unscattered electrons for the gas-filled WGTS $\dot{N}_{\textrm{e}}(\mathcal{N})$ with the rate for the evacuated WGTS $\dot{N}_{\textrm{e}}(0)$ gives the zero-scattering probability $P_0$ as (see eq.~\eqref{eq:scattering_prob})
\begin{equation}
\dot{N}_{\textrm{e}}(\mathcal{N})=P_0(\mathcal{N})\cdot\dot{N}_{\textrm{e}}(0).
\end{equation}
Now $\mathcal{N} \cdot \sigma$ can be determined by plugging $P_0$ into eq.~(\ref{eq:scattering_prob}).

To reach an appropriate precision of \num{0.1}\% for $\dot{N}_{\textrm{e}}(\mathcal{N})$ at nominal column density of \SI{5e21}{\per\metre\squared}, a measurement time of \SI{2.5}{\minute} is required for an e-gun rate of $\SI{1e5}{electrons\per\second}$. Rate instabilities directly translate to uncertainties in scattering probability which means a rate stability of the order of \num{0.1}\% needs to be achieved during the whole e-gun measurement cycle\footnote{This means \num{0.1}\% stability from beginning of the measurement with empty WGTS through filling of the WGTS until the end of the measurement with filled WGTS}. 

Using $\frac{\Delta(\mathcal{N}\cdot\sigma)}{\mathcal{N}\cdot \sigma}=\frac{\Delta P_0}{P_0}\frac{1}{\ln P_0}$, with $P_0\approx\num{18.3}\%$ for e-gun electrons at the tritium endpoint, the e-gun stability specification implies a relative $\mathcal{N}\cdot \sigma$ uncertainty of \num{6e-4}.	Furthermore, the finite angular resolution of the beam as well as drifts of the angle between magnetic field lines and e-gun beam need to be considered.  
This results in a relative e-gun measurement uncertainty  $\left(\frac{\Delta(\mathcal{N} \cdot \sigma)}{\mathcal{N}\cdot \sigma}\right)_\textrm{abs}$  on the absolute value of $\mathcal{N}\cdot \sigma$ of about \num{0.15}\% for the given e-gun specifications. It should be noted that the described e-gun measurement requires an evacuated WGTS, so it temporarily blocks neutrino-mass data-taking. 
	    
\subsection{Determination of changes in {\normalfont $\mathcal{N}$}} \label{subsec:relative-uncertainties}
In between the measurements described in sec.~\ref{subsec:egun_measurement}, operational source parameters can vary, causing column density changes. Since the inelastic scattering cross section is constant over time, this means column density fluctuations $\left(\frac{\Delta \mathcal{N}}{\mathcal{N}}\right)_\textrm{rel}$ need to be covered by the \num{0.2}\% uncertainty budget:
\begin{equation} 
	\label{eq:rhod_rel_budget}
	\frac{\Delta(\mathcal{N}\cdot\sigma)}{\mathcal{N}\cdot\sigma} = \left(\frac{\Delta \mathcal{N}}{\mathcal{N}}\right)_\textrm{rel}\le\num{2e-3}.
\end{equation}

Operational parameters that influence the column density are the pressure in the pressure-controlled buffer vessel $p_{\textrm{B}}$ determining the WGTS injection pressure, the WGTS temperature $T$ and the TMP pumping speed that is related to the WGTS beam tube outlet pressure $p_\textrm{ex}$. Eq.~\eqref{eq:rhod_rel_budget} can be used to obtain a limit for the column density fluctuation $\frac{\Delta \mathcal{N}_x}{\mathcal{N}}$ of each of the 3 mentioned operational parameters $x$.

Assuming all  mentioned operational parameters are uncorrelated and considering additional contributions from variations of the tritium purity (affecting the \textbeta-decay electron rate stability), the following requirement needs to hold for each parameter $x$:
\begin{equation}
\label{eq:rel_rhod_max}
\frac{\Delta \mathcal{N}_x}{\mathcal{N}}=\frac{1}{\sqrt{4}}\left(\frac{\Delta \mathcal{N}}{\mathcal{N}}\right)_\textrm{rel} \le \num{1e-3}.
\end{equation}
Therefrom stability requirements for the monitored parameters can be derived.
The impact of a changing buffer vessel pressure $p_{\textrm{B}}$ on $\mathcal{N}$ is calculated using its influence on the throughput $q$. Both are linked through the conductance $C$ 
of the tube system connecting buffer vessel and beam tube inlet (at injection pressure $p_\textrm{in}$) with:
\begin{equation}\label{eq:C_sum}
\frac{1}{C}=\frac{1}{C_{\textrm{capillary}}}+\frac{1}{C_{\textrm{loop}}},
\end{equation}
\begin{align}
q&=(p_{\textrm{B}}-p_{\textrm{in}})\cdot C, \quad \textrm{\quad for $T$=const. and $ p_{\textrm{in}}\ll p_{\textrm{B}}$:}\\
\frac{\Delta q}{q}&\approx\underbrace{\left(C+p_{\textrm{B}}\cdot\frac{\delta C}{\delta p_{\textrm{B}}}\right)\frac{p_{\textrm{B}}}{q}}_{\beta_{p_{\textrm{B}}}}\frac{\Delta p_{\textrm{B}}}{p_{\textrm{B}}} \label{eq:dq_q_pB}.
\end{align} 
All conductance values except that of the capillary $C_\textrm{capillary}$ feeding gas into the injection chamber can be neglected. The feed capillary is thermally coupled to the WGTS cooling system and therefore stabilised at \SI{30}{\kelvin}. Since the flow through the capillary is hydrodynamic, the throughput can be calculated using  Poiseuille's formula~\cite{Landau1989}.
The average pressure $\bar{p}$ can be approximated with $p_{\textrm{B}}/2$, since $p_\textrm{in}\ll p_\textrm{B}$. Using the conductance of a tube, eq.~\eqref{eq:dq_q_pB}, $p_{\textrm{B}}=\SI{10}{\milli\bar}$ and ${q=\SI{1.8}{\milli\bar\litre\per\second}\cdot\frac{\SI{30}{\kelvin}}{\SI{273.15}{\kelvin}}}$, we obtain $\beta_{p_{\textrm{B}}}\approx \num{2}$. 

\begin{figure}
  \subfigure[]{\includegraphics[width=0.49\textwidth]{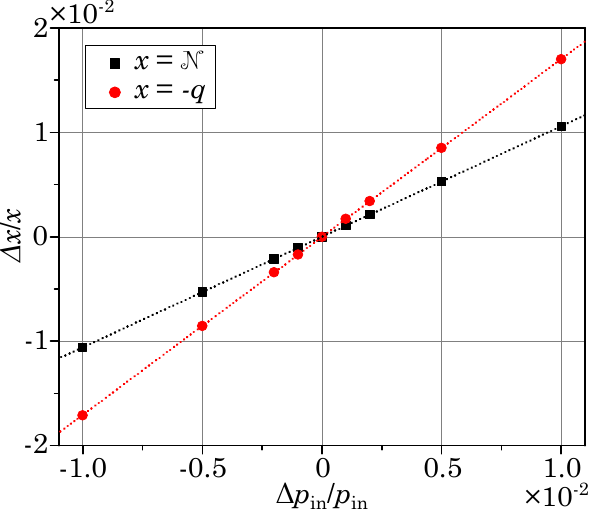}\label{fig:dN_dq}} \quad	     \subfigure[]{\includegraphics[width=0.49\textwidth]{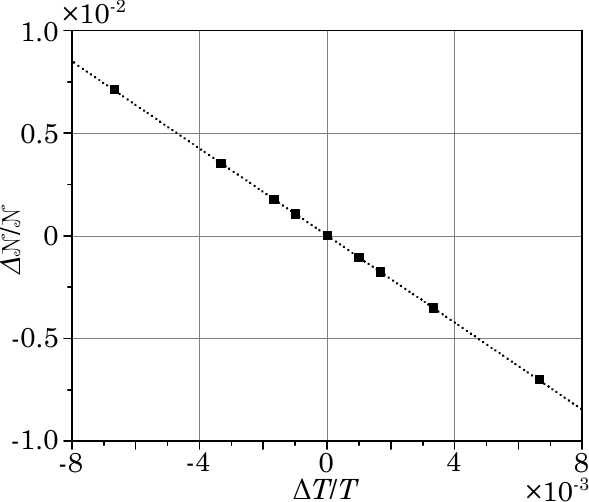}\label{fig:dN_dT}}
  \caption{a) Influence of injection pressure $p_{\textrm{in}}$ on column density ($\mathcal{N}$) and throughput ($q$). Using the slope of the linear regression (dotted lines) of both data sets, $\alpha_{p_{\textrm{in}}}$ and  $\left(\beta_{p_{\textrm{in}}}\right)^{-1}$ can be obtained. b)  WGTS column density dependence on relative temperature change for a mean temperature of \SI{30}{\kelvin}. The coefficient $\alpha_T$ is obtained as -1.06 from the slope of the linear regression (dotted line). }
\end{figure}

The calculation of the conductance is taken to be as accurate as \num{10}\%, which matches the quality that can be expected in a dedicated measurement. This translates to a $\beta_{p_{\textrm{B}}}$ uncertainty of \num{10}\%, as well. Column density and throughput variations are related through
\begin{align}
\frac{\Delta\mathcal{N}}{\mathcal{N}}&=\frac{d\mathcal{N}}{d q}\cdot\frac{\Delta q}{\mathcal{N}}=\underbrace{\frac{d\mathcal{N}}{d q}\cdot\frac{q}{\mathcal{N}}}_{\alpha_q}\cdot \frac{\Delta q}{q}. \label{eq:dN_dq}\\
\intertext{For constant temperature this can be rewritten as}
\frac{\Delta\mathcal{N}}{\mathcal{N}}&=\underbrace{\frac{\delta \mathcal{N}}{\delta p_{\textrm{in}}}\cdot\frac{p_{\textrm{in}}}{\mathcal{N}}}_{\alpha_{p_{\textrm{in}}}}\cdot \left(\underbrace{\frac{\delta q}{\delta p_{\textrm{in}}}\cdot\frac{p_{\textrm{in}}}{q}}_{\beta_{p_{\textrm{in}}}}\right)^{-1}\frac{\Delta q}{q}. \label{eq:dN_dq_dp}
\end{align} 
Thus, the coefficients are related as follows:  $\alpha_q=\alpha_{p_{\textrm{in}}}\left(\beta_{p_{\textrm{in}}}\right)^{-1}$ with $\frac{\Delta \mathcal{N}}{\mathcal{N}}=\alpha_{p_{\textrm{in}}}\frac{\Delta p_{\textrm{in}}}{p_{\textrm{in}}}$, $\frac{\Delta q}{q}=\left(\beta_{p_{\textrm{in}}}\right)^{-1}\frac{\Delta p_{\textrm{in}}}{p_{\textrm{in}}}$. The values can be read off the slopes in fig.~\ref{fig:dN_dq}:  $\alpha_{p_{\textrm{in}}}\approx \num{1.06}$, $\beta_{p_{\textrm{in}}}\approx \num{1.7}$ and thus $\alpha_q\approx \num{0.62}$.
This allows us to compute  column density changes that are induced by buffer vessel pressure fluctuations:
\begin{equation}
\label{eq:drod_pb}
\frac{\Delta \mathcal{N}}{\mathcal{N}}=\alpha_q \cdot\beta_{p_{\textrm{B}}}\cdot\frac{\Delta  p_{\textrm{B}}}{p_{\textrm{B}}} \approx 1.24\frac{\Delta  p_{\textrm{B}}}{p_{\textrm{B}}}.
\end{equation}
If the column density is not corrected for, eq.~\eqref{eq:drod_pb} and~\eqref{eq:rel_rhod_max} imply a buffer vessel pressure stability requirement of
\num{8e-4}. The pressure stability reached by Priester et al.~\cite{Priester2015} in test operation of the gas circulation was better than \num{2e-4} and thus well within this limit.

The influence of temperature fluctuations on column density is derived by using different temperature values for the WGTS beam tube gas density calculation with all other input parameters fixed. The result is depicted in fig.~\ref{fig:dN_dT}. Therefrom, a coefficient $\alpha_T\approx\num{-1.06}$ with $\frac{\Delta \mathcal{N}}{\mathcal{N}}=\alpha_T\cdot\frac{\Delta T}{T}$ is derived. This implies a relative temperature stability requirement of \num{9e-4} between the e-gun column density measurements.

Test measurements showed the achievable beam tube temperature stability for one week of operation, which is significantly longer than the planned e-gun measurement time steps, to be well within this limit~\cite{Grohmann2013}, which was confirmed by measurements with the full \katrin~beamline~\cite{FirstLightKrypton2018}.

If operational parameter variations larger than $\num{1e-3}$ occur during neutrino-mass measurements, the gas model described in the previous section can be used to update the column density value $\mathcal{N}$. For example, pressure readings can be used to account for changes of the pressure in the buffer vessel $\Delta p_B$. This procedure even allows reduction of the systematic uncertainty from eq.~\eqref{eq:rhod_rel_budget} related to column density fluctuations $\left(\frac{\Delta \mathcal{N}}{\mathcal{N}}\right)_\textrm{rel}$. 

The limits of the described compensation are mainly determined by the accuracy of the gas model calculation. Absolute values of $\mathcal{N}(p_\textrm{in})$ or equally $p_\textrm{in}(\mathcal{N})$ can be calculated with an uncertainty of \num{2}\%, as shown in section~\ref{subsec:egun_measurement}. It still needs to be evaluated how accurately changes in column density $\Delta \mathcal{N}_x$ caused by variations of source parameter $x$ can be calculated. Therefore, $\Delta \mathcal{N}_x$ is computed for two inlet pressures that deviate by \num{5}\%. Thus, the relative modelling uncertainty $\left(\frac{\Delta \mathcal{N}_\textrm{x}}{\mathcal{N}}\right)_\textrm{m}$ reads
\begin{align}
\label{eq:rel_rhod_pin}
\left(\frac{\Delta \mathcal{N}_x}{\mathcal{N}}\right)_\textrm{m}=&\frac{\mathcal{N}\left(p_{\textrm{in}},x\right)-\mathcal{N}\left(p_{\textrm{in}},x\left(\frac{\Delta x}{x}+1\right)\right)-\mathcal{N}\left(\num{1.05}\cdot p_{\textrm{in}},x\right)} {\mathcal{N}\left(p_{\textrm{in}},x\right)}\nonumber \\
& \frac{+\mathcal{N}\left(\num{1.05}\cdot p_{\textrm{in}},x\left(\frac{\Delta x}{x}+1\right)\right)}{\mathcal{N}\left(p_{\textrm{in}},x\right)}.
\end{align}
Moreover, the variations of the model input parameters $p_\textrm{in}$ and $p_\textrm{ex}$ can only be derived from the monitored values of buffer vessel pressure and pressure next to the TMP in the first WGTS pump port.
Thus, the uncertainty of this calculation, about \num{10}\% for $\frac{\Delta p_\textrm{in}}{p_\textrm{in}}$ and \num{40}\% for $\frac{\Delta p_\textrm{ex}}{p_\textrm{ex}}$~\cite{Kuckert2016}, needs to be considered in addition to eq.~\eqref{eq:rel_rhod_pin}. 

\begin{figure}
     \subfigure[]{\includegraphics[width=0.49\textwidth]{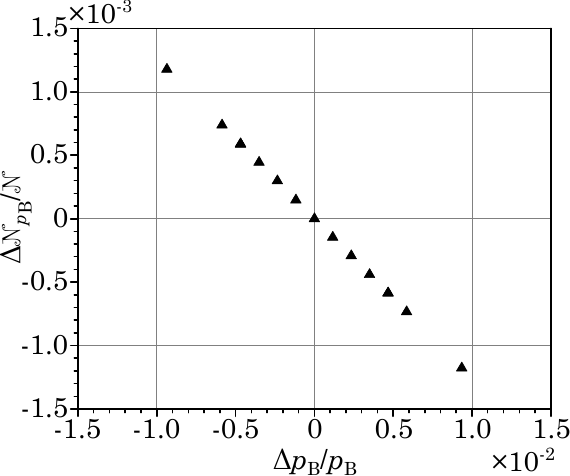}\label{fig:relrhod_pin}}\quad
     \subfigure[]{\includegraphics[width=0.49\textwidth]{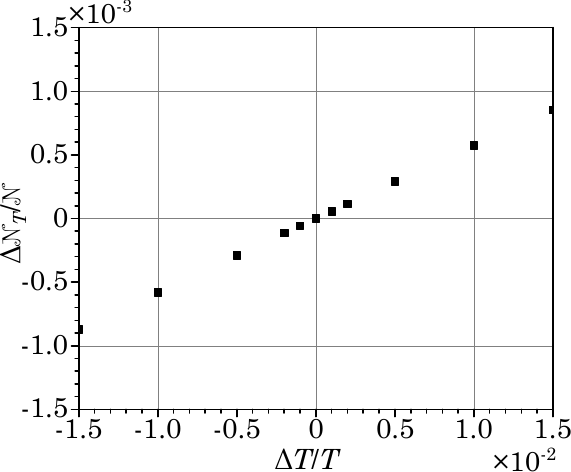}\label{fig:relrhod_T}}
   \caption{Uncertainty of calculated column density changes $\Delta \mathcal{N}_x$ for variations of  buffer vessel pressure ($x=p_\textrm{B}$) and beam tube temperature ($x=T$).
   The  limit from equation~\eqref{eq:rel_rhod_max} is added within the dashed red lines.}
\end{figure}

Adding the different contributions, the relative calculation uncertainty of $\Delta \mathcal{N}_x$ can be derived as shown in fig.~\ref{fig:relrhod_pin} and fig.~\ref{fig:relrhod_T}. If changes in parameters $x$ are accounted for in the gas modelling, this implies that the requirements on buffer vessel pressure and beam tube temperature stability can be relaxed to \num{8e-3} and \num{1.8e-2} respectively (compare eq.~\eqref{eq:rel_rhod_max}), while still matching the \num{2e-3} accuracy requirement on $\mathcal{N} \cdot \sigma$.

If temperature and inlet pressure are stable on the \num{1e-3} level as required~\cite{KATRIN2004}, column density changes can be modelled with an accuracy better than \num{3e-4}. 
Inserting the e-gun measurement accuracy of \num{1.5e-3} as stated above, the total  uncertainty on $\mathcal{N} \cdot \sigma$ is thus even below \num{1.6e-3} which reduces the related systematic neutrino mass uncertainty as discussed in the following section.

\subsection{Implications of gas dynamics uncertainties for the neutrino mass analysis}  
\label{subsec:delta_mnu}
Any unaccounted effect that modifies the electron energy spectrum introduces a systematic shift in the measured neutrino mass squared~\cite{Robertson1988}.

With regard to the description of gas dynamics, the product of column density $\mathcal{N}$ and scattering cross section $\sigma$ is a first-order effect: it is the property that 
has the largest influence on the electron spectrum as it determines the (average) scattering probabilities. 

Second-order effects require the consideration of the gas density distribution and detailed knowledge of the column density and the scattering cross section.
Such second-order effects are basically caused by the inhomogeneous bulk gas velocity distribution as well as by inhomogeneities of the magnetic field, temperature and electrostatic potential. 

To investigate the various gas-model-related systematic effects on the neutrino mass measurement, the method of ensemble testing is used~\cite{Kleesiek2014,Kleesiek:2018mel}. For each analysis 5000 toy \katrin~measurement spectra are generated using the source spectrum calculation (SSC) package that is implemented in the \katrin~simulation software~\cite{Kleesiek2014,Kleesiek:2018mel}. In this generation of electron spectra the neutrino mass is assumed to be zero ($m_{\upnu_0}=0$) for the sake of simplicity and without loss of generality. Reasonable values are also chosen for the other undetermined parameters of the spectrum: its endpoint, amplitude and the background rate. Statistical randomness of a measurement is implemented using Poisson fluctuations of the derived count rates.

In a second step, an analytical spectrum is calculated in a similar procedure but without statistical fluctuations. It includes the systematic effect to be analysed (e.g. shift of a gas model parameter towards lower value). The free parameters of the analytical spectrum, among those the neutrino mass squared, are determined through a fit of the analytical spectrum to the generated toy data~\cite{Kleesiek2014,Kleesiek:2018mel}.
Therefore, the negative log-likelihood is minimised and a best fit value for the neutrino mass squared, $m_{\upnu\textrm{fit}}^2$, is derived. 

\begin{figure}
    \centering
    \includegraphics[width=0.9\textwidth]{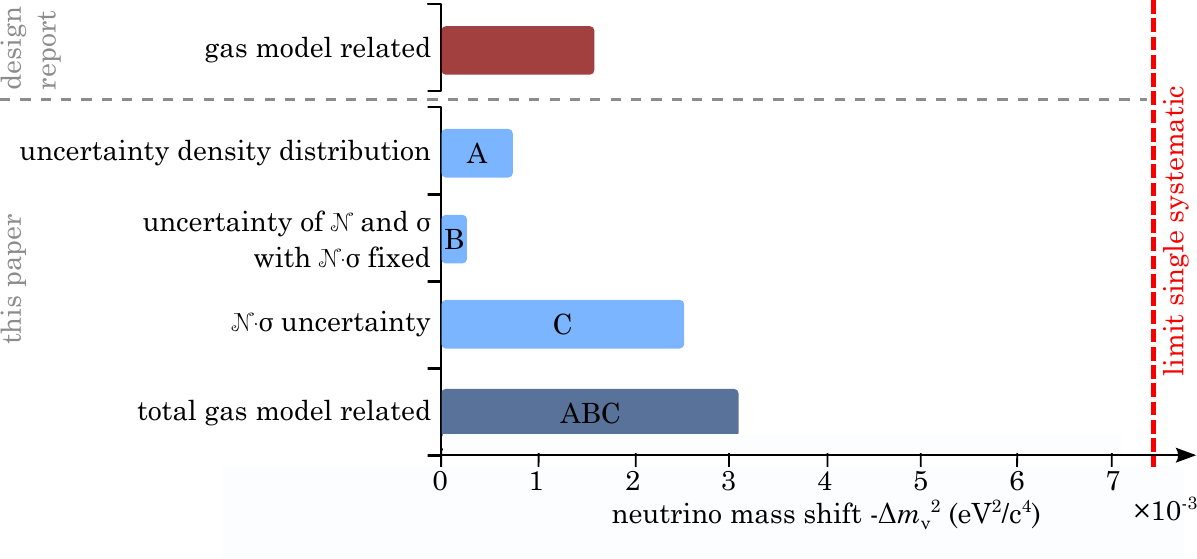}
    \caption{Summary of gas-model-related uncertainties obtained in this paper (denoted by A, B and C) and in the \katrin~design report~\cite{KATRIN2004} and their induced systematic neutrino mass shift $\Delta m_\upnu^2$. Including second-order effects, the revised overall gas-related uncertainty (ABC) is calculated respecting all uncertainties A, B and C at once in one single ensemble test. The obtained uncertainty is almost twice as large as assumed in ref.~\cite{KATRIN2004}. Compared to the limit for a single systematic effect ($7.5\cdot\SI{e-3}{\electronvolt\squared}/\textrm{c}^4$), however, the impact of gas model uncertainties is still moderate.}
    \label{fig:mnu_effect}
\end{figure}

Pursuing this procedure for an ensemble of generated spectra, the systematic neutrino mass squared shift $\Delta m_\upnu^2$ that is induced by the analysed effect can be determined
using the  mean $\mu(m_{\upnu\textrm{fit}}^2)$ of the obtained  $m_{\upnu\textrm{fit}}^2$ distribution~\cite{Kleesiek2014,Kleesiek:2018mel}
\begin{equation}
	\Delta m_\upnu^2=\mu(m_{\upnu\textrm{fit}}^2)-m_{\upnu0}^2.
\end{equation}   
\begin{itemize}
	\item The impact of first-order gas dynamical effects, e.g. of the accuracy of the parameter $\mathcal{N} \cdot \sigma$, is investigated by introducing relative $\mathcal{N}\cdot\sigma$ differences of 0.2\% for the two spectra used in the analysis. All other experimental parameters (analysis window, background, \dots) were chosen according to the standard settings defined in ref.~\cite{KATRIN2004}.
	This produces a neutrino mass squared shift (C in fig.~\ref{fig:mnu_effect}) of
	\begin{align}
	\left.\Delta m_\upnu^2\right|_\text{C}=\left( -2.62\pm 0.25\right)\cdot\SI{e-3}{\electronvolt\squared}/\textrm{c}^4. \nonumber
	\end{align}
	
	\item The second-order effect of a limited column density accuracy of 2\%, while $\mathcal{N} \cdot \sigma$ is assumed to be known precisely, is quantified to be (B in fig.~\ref{fig:mnu_effect})
	\begin{equation}
		\left.\Delta m_\upnu^2\right|_\text{B}=\left(-0.26\pm 0.25\right)\cdot\SI{e-3}{\electronvolt\squared}/\textrm{c}^4. \nonumber
	\end{equation}
	by using a similar procedure. 
	
	\item The impact of the accuracy of the actual density profile for a fixed column density is tested by using two density profiles deviating by up to 5\% . Here the ensemble testing yields a systematic neutrino mass squared shift of (A in fig.~\ref{fig:mnu_effect})
	\begin{equation}
		\left.\Delta m_\upnu^2\right|_\text{A}=\left(-0.75\pm 0.24\right)\cdot\SI{e-3}{\electronvolt\squared}/\textrm{c}^4. \nonumber
	\end{equation} 
\end{itemize}
Since the mentioned systematic effects are correlated, they need to be combined  in a single ensemble test. Doing so (for $\Delta\mathcal{N} \cdot \sigma/(\mathcal{N}\cdot\sigma)=\num{0.2}\%$) results in a systematic neutrino mass shift of (ABC in fig.~\ref{fig:mnu_effect})
\begin{equation}
   \left.\Delta m_\upnu^2\right|_\text{ABC}=\left(-3.06\pm 0.24\right)\cdot\SI{e-3}{\electronvolt\squared}/\textrm{c}^4.  \nonumber
\end{equation} 
This value  represents the total systematic uncertainty related to the description of gas dynamical processes in the source and transport section of \katrin. As depicted in fig.~\ref{fig:mnu_effect} it is almost twice as large as the gas-related effect assumed in ref.~\cite{KATRIN2004}. In previous analyses only primary gas model effects, e.g. the uncertainty of $\mathcal{N}\cdot \sigma$, had been considered.
However, the revised overall gas-related uncertainty does not constitute a dominant neutrino mass shift. It is still less than half of the limiting value for a single systematic effect (compare fig.~\ref{fig:mnu_effect}). 

\section{Conclusion}\label{sec:conclusion}
\katrin~relies on proper modelling of the spectrum of electrons stemming from tritium \textbeta-decay which implies an accurate knowledge of the transport processes in the  source. One of the major systematic effects of a gaseous source type as used in \katrin~is the description of the inelastic electron-gas molecule scattering process. 
Being closely linked to the gas flow in the source section it  underlines the importance of the description of gas dynamics for the modelling of the source electron spectra and thus for the \katrin~sensitivity.

In this paper we presented several gas flow calculations of different domains constituting the \katrin~source section over a wide range of gas rarefaction. Those calculations were put together to form an intricate source gas model to be used in the neutrino mass analysis. Together with the input from regular calibration measurements and continuous monitoring of source operational parameters this model allows an accurate online modelling of the gas density and velocity distributions, which is also an important input for plasma simulations.

To analyse the impact of the modelling of gas dynamics on the neutrino mass measurement, different gas-related systematic uncertainties  were considered based on a realistic source model. It was shown that the related systematic uncertainty of $\Delta m_\upnu^2=\left(-3.06\pm 0.24\right)\cdot\SI{e-3}{\electronvolt\squared}/\textrm{c}^4$ is within the allowed budget. 

This demonstrates that gas dynamical processes in the source are well understood and that the described gas model in combination with regular column-density calibration measurements with an electron gun can be used in the calculation of electron spectra for the actual neutrino mass measurement. Experimental verification of the presented model is currently scheduled as part of the final stages of \katrin~commissioning.
\section*{Acknowledgements}
We acknowledge the support of the Helmholtz Association (HGF), the German Ministry for Education and Research BMBF (05A17PM3, 05A17PX3, 05A17VK2, and 05A17WO3), the Helmholtz Alliance for Astroparticle Physics (HAP), the Helmholtz Young Investigator Group VH-NG-1055, and the DFG graduate school KSETA (GSC~1085). We are grateful to D. S. Parno and M. Schl\"osser for very valuable comments and discussions.

{\small
	%% if using biblatex
%	\printbibliography
	
	%% if using bibtex
	\bibliographystyle{IEEEtran} % abbrv, unsrt, ieeetr
	\bibliography{ms}
}

\end{document}